\documentclass[draft,prd,preprint,tightenlines,floatfix,showpacs,preprintnumbers,nofootinbib,eqsecnum]{revtex4}
 \usepackage[dvips,final]{graphicx}
  \usepackage{amssymb}
   \usepackage{amsmath}
    \usepackage{epsfig}
     \usepackage{bm}
      \usepackage{pifont}

\def\muF{\relax\ifmmode\mu_\text{F}^2\else{$\mu_\text{F}^2${ }}\fi}
\def\muR{\relax\ifmmode\mu_\text{R}^2\else{$\mu_\text{R}^2${ }}\fi}
\def\muO{\relax\ifmmode{\mu_{0}^{2}}\else{$\mu_{0}^{2}${ }}\fi}
\def\Mev{\relax\ifmmode{\text{MeV}}\else{MeV{ }}\fi}
\def\x{\overline{x} \,}
\def\y{\overline{y} \,}

\def\MS{$\overline{\text{MS}\vphantom{^1}}${ }}

\def\Li{\relax\ifmmode{\textbf{Li}_{2}}\else{Li$_2${ }}\fi}
\def\Im{\relax{\textbf{Im}{}}}

\newcommand{\gev}[1]{\relax\ifmmode{\text{GeV}^{#1}}\else{GeV$^{#1}${ }}\fi}

\def\asb{\relax\ifmmode \bar{\alpha}_s\else{$ \bar{\alpha}_s${ }}\fi}
\def\as{\relax\ifmmode \alpha_s\else{$ \alpha_s${ }}\fi}
\def\acal{\relax\ifmmode{\cal A}\else{${\cal A}${ }}\fi}
\newcommand\convo[1]{\mathop{\otimes}\limits_{#1}}

\begin{document}
\thispagestyle{empty}
\date{\today}
\preprint{\hbox{UA/NPPS-02-05, RUB-TPII-04/05, JINR-E2-2005-155}}

\title{Analyticity properties of three-point functions in QCD
       beyond leading order\\}
\author{A.~P.~Bakulev}
 \email{bakulev@theor.jinr.ru}
  \affiliation{Bogoliubov Laboratory of Theoretical Physics, JINR,
   141980 Dubna, Russia\\}

\author{A.~I.~Karanikas}
 \email{akaranik@cc.uoa.gr}
  \affiliation{University of Athens, Department of Physics,
               Nuclear and Particle Physics Section,
               Panepistimiopolis, GR-15771 Athens, Greece\\}

\author{N.~G.~Stefanis}
 \email{stefanis@tp2.ruhr-uni-bochum.de}
  \affiliation{Institut f\"{u}r Theoretische Physik II,
               Ruhr-Universit\"{a}t Bochum,
               D-44780 Bochum, Germany}

\begin{abstract}
The removal of unphysical singularities in the perturbatively
calculable part of the pion form factor---a classic example
of a three-point function in QCD---is discussed.
Different ``analytization'' procedures in the sense of Shirkov
and Solovtsov are examined in comparison with standard QCD
perturbation theory.
We show that demanding the analyticity of the partonic amplitude
as a \emph{whole}, as proposed before by Karanikas and Stefanis,
one can make infrared finite not only the strong running coupling
and its powers, but also cure potentially large logarithms (that
first appear at next-to-leading order) containing the factorization
scale and modifying the discontinuity across the cut along the
negative real axis.
The scheme used here generalizes the Analytic Perturbation Theory
of Shirkov and Solovtsov to non-integer powers of the strong
coupling and diminishes the dependence of QCD hadronic quantities
on all perturbative scheme and scale-setting parameters, including
the factorization scale.
\end{abstract}
\maketitle


\cleardoublepage
\section{Introduction}
\label{sec:intro}

The phenomenology of QCD exclusive processes depends in a crucial way
on the analytic properties of hadronic (hard) scattering amplitudes as
functions of the strong running coupling.
A perturbatively calculable short-distance part of the reaction
amplitude at the parton level is isolated either by subtraction or by
factorization.
To get a quantitative interpretation of such quantities in practice
and compare them with experimental data, one has to get rid of the
artificial Landau singularity at
$Q^2=\Lambda^{2}_\text{QCD}$ ($\Lambda_\text{QCD}\equiv \Lambda$
in the following), where $Q^2$ is the large mass scale in the process.
A proposal to solve this problem (in the spacelike region) without
introducing exogenous infrared (IR) regulators, like an effective, or
a dynamically generated, gluon mass \cite{Cor82} (see, for instance,
\cite{PP79,JSL87,Ste89,JiAm90,SB93D,MS93,BJPR98,Mul98,Ste99} for such
applications), was made by Shirkov and Solovtsov (SS)
\cite{SS97,Shi98,SS99}, based on general principles of local Quantum
Field Theory.
This theoretical framework---termed Analytic Perturbation Theory
(APT)---was further expanded beyond the one-loop level of two-point
functions to define an analytic\footnote{The term `analyticity' is
used here as a synonym for `spectrality' and `causality'
\cite{Shi98}.} coupling and its powers in the timelike region
\cite{SS98,DVS00unp,Shi00,SS01,KM01,Mag99,BMMR02,Ale05},
embracing previous attempts \cite{Rad82,KP82,BB94,BBB95,MS97,BRS00} in
this direction.\footnote{A somewhat different approach was reviewed
recently in \cite{Nes03}; see also \cite{NP04}.}

However, first applications \cite{SSK99,SSK00} of this sort of approach
to three-point functions, beyond the leading order of QCD perturbation
theory, have made it clear that, ultimately, there must be an extension
of this formalism from the level of the running coupling and its powers
to the level of amplitudes.
The reason is that in three-point functions at the next-to-leading
order (NLO) level, and beyond, logarithms of a distinct scale
(serving as the factorization or evolution scale) appear
that though they do not change the nature of the Landau pole, they
affect the discontinuity across the cut along the negative real axis
$-\infty < Q^2 < 0$.
On account of factorization, we expect that this effect should be
small, of the order of a few percent, because any change caused by the
variation of the factorization scale should be of the next higher
order.
However, to achieve a high-precision theoretical prediction, one should
reduce this uncertainty, lifting the limitations imposed by the lack of
knowledge about uncalculated higher-order corrections.
To encompass such logarithmic terms in the ``analytization'' procedure,
one should demand the analyticity of the partonic amplitude as a
\emph{whole} \cite{KS01,Ste02} and calculate the dispersive image of
the coupling (or of its powers) \emph{in conjunction} with these
logarithms.
This Karanikas--Stefanis (KS) ``analytization'' scheme effectively
amounts to the generalization of APT to non-integer powers of the
running coupling: Fractional APT (FAPT), as we shall show below.

In this work we expand the Shirkov--Solovtsov ``analytization''
approach to include the dispersive images of such terms, using as a
case study the pion form factor at NLO in the \MS scheme with various
renormalization-scale settings and also in the $\alpha_V$-scheme
\cite{BrodskyL95}.
To this end, we contrast the KS ``analytization'' with the \emph{naive}
\cite{SSK99,SSK00} and the \emph{maximal} \cite{BPSS04}
``analytization'' procedures and work out their key mutual differences
as they first appear in NLO, while a fully-fledged analysis of FAPT\
is given in an accompanying paper \cite{BMS05}.
We argue that augmenting the \MS scheme with the KS ``analytization''
prescription provides an optimized method to calculate perturbatively
higher-order corrections to partonic ``observables'' in QCD because it
practically eliminates all scheme and scale-setting ambiguities owing
to the renormalization and factorization scales.
It is worth emphasizing at this point that the focus of Ref.\
\cite{KS01} was on the calculation of power corrections to the pion's
electromagnetic form factor.
Such contributions are outside the scope of the present investigation.

The plan of this paper is as follows.
In Sec.\ \ref{sec:standardQCD} we review the convolution formalism for
the calculation of the short-distance part of the pion form factor
within perturbative QCD at NLO.
In Sec.\ \ref{sec:analyticity} we discuss the Shirkov--Solovtsov type
``analytization'' procedures \cite{SS97,SSK99,SSK00,KS01,Ste02,BPSS04}
and work out their mutual differences, focusing on the KS
``analytization'' and its properties.
This discussion extends and generalizes the original KS analysis
that covered only the LO of the perturbative expansion of the pion
form factor and ignoring evolution.
Section \ref{sec:FFanalytic} contains the results for the factorized
pion form factor in different schemes and with different scale
settings, employing the KS ``analytization'' in comparison with those
based on APT and also standard QCD perturbation theory in NLO.
Our conclusions with a summary of our main results are presented in
Sec.\ \ref{sec:concl}.
Some important technical details are collected in three appendices.

\section{Factorizable Part of the Pion Form Factor at
         NLO in Standard QCD Perturbation Theory}
\label{sec:standardQCD}

The leading-twist factorizable part of the electromagnetic pion form
factor can be expressed as a convolution in the form \cite{ER80,LB80}
\begin{eqnarray}
 F_{\pi}^\text{Fact}(Q^{2}; \muR)
=
   \Phi_{\pi}^{*}(x,\muF)\convo{x}
     T_\text{H}(x,y,Q^{2};\muF,\muR)\convo{y}
      \Phi_{\pi}(y,\muF)\, ,
\label{eq:pff-Fact}
\end{eqnarray}
where $\otimes$ denotes the usual convolution symbol
($A(z)\convo{z}B(z) \equiv \int_0^1 dz A(z) B(z)$)
over the longitudinal momentum fraction variable $x$ ($y$) and
$\mu_\text{F}$ represents the factorization scale at which the
separation between the long- (small transverse momentum) and
short-distance (large transverse momentum) dynamics takes place,
with $\mu_\text{R}$ labelling the renormalization (coupling
constant) scale.
The nonperturbative input is encoded in the pion distribution
amplitude (DA) $\Phi_{\pi}(y,\muF)$, whereas the short-distance
interactions are represented by the hard-scattering amplitude
$T_\text{H}(x,y,Q^{2};\muF,\muR)$.
This is the amplitude for a collinear valence quark-antiquark pair
with total momentum $P$ struck by a virtual photon with momentum
$q$, satisfying $q^2=-Q^2$,
to end up again in a configuration of a parallel valence
quark-antiquark pair with momentum $P'=P+q$.
It can be calculated perturbatively in the form of a power-series
expansion in the QCD coupling, the latter to be evaluated at the
reference scale of renormalization $\muR$:
\begin{eqnarray}
   T^\text{NLO}_\text{H}(x,y,Q^2;\muF,\muR) =
       \alpha_s\left(\muR\right)\,  T_\text{H}^{(0)}(x,y,Q^2)
         + \frac{\alpha_s^2\left(\muR\right)}{4 \pi} \,
             T_\text{H}^{(1)}(x, y, Q^2;\muF, \muR)\, .
\label{eq:TH}
\end{eqnarray}
The leading-order (LO) contribution to $T_\text{H}(x, y, Q^2;\muF)$
reads
\begin{eqnarray}
   T_\text{H}^{(0)}(x,y,Q^2)
   = \frac{N_\text{T}}{Q^2} \,\frac{1}{\x \y}
   \equiv
     \frac{1}{Q^2}\, t_\text{H}^{(0)}(x,y)\,,
\label{eq:THLOpff}
\end{eqnarray}
where
\begin{eqnarray}
   N_\text{T} = \frac{2\, \pi \, C_\text{F}}{C_\text{A}}
       = \frac{8\pi}{9}\, ,
\label{eq:NTpff}
\end{eqnarray}
$C_\text{F}=\left(N_\text{c}^{2}-1\right)/2N_\text{c}=4/3$,
$C_\text{A}=N_\text{c}=3$ are the color factors of
$SU(3)_\text{c}$, and the notation $\bar{z} \equiv 1-z$ has been used.
The usual color decomposition of the NLO correction
\cite{MNP98}---marked by self-explainable labels---is given by
(omitting the variables $x$ and $y$)
\begin{eqnarray}
 Q^2T_\text{H}^{(1)}\left(Q^2;\muF,\muR\right)
  = C_\text{F}\,t_\text{H}^{(1,\text{F})}
  \left(\frac{\muF}{Q^2}\right)
  + b_0\,t_\text{H}^{(1,\beta)}\left(\frac{\muR}{Q^2}\right)
  + C_\text{G}\,t_\text{H}^{(1,\text{G})}\,,
\label{eq:THNLOpff}
\end{eqnarray}
where $C_\text{G}=(C_\text{F}-C_\text{A}/2)$
and $b_0$ is the first coefficient of the $\beta$ function,
see Appendix \ref{app:QCD-PT}, Eq.\ (\ref{eq:beta0&1}).
Here we explicitly factorized out a trivial $1/Q^2$ dependence
and used for the coefficients in front of each factor the
notation $t_\text{H}$ with appropriate superscripts.

With reference to the application of the Brodsky--Lepage--Mackenzie
(BLM)\ \cite{BLM83} scale setting in fixing the renormalization point
later on, we single out the $b_0$-proportional (i.e., the
$N_f$-dependent) term, given by
\begin{subequations}\label{eq:TH1beta}
\begin{eqnarray}
  t_\text{H}^{(1,\beta)}\left(x,y;\frac{\muR}{Q^2}\right)
   &=& t_{\text{H},1}^{(1,\beta)}\left(x,y\right)
    +  t_{\text{H},2}^{(1,\beta)}\left(x,y;\frac{\muR}{Q^2}\right)
\label{eq:TH1beta-sum}
\end{eqnarray}
with
\begin{eqnarray}
 t_{\text{H},1}^{(1,\beta)}\left(x,y\right)
   &=& t_\text{H}^{(0)}\left(x,y\right)\,
        \left[\frac{5}{3} - \ln (\x \y)\right]
\label{eq:TH11beta}
 \\ [1.5mm]
  t_{\text{H},2}^{(1,\beta)}\left(x,y;\frac{\muR}{Q^2}\right)
   &=& t_\text{H}^{(0)}\left(x,y\right)\,
       \ln \frac{\muR}{Q^2}\, ,
\label{eq:TH12beta}
\end{eqnarray}
\end{subequations}
and present the color singlet part of $t_\text{H}$
in the form
\begin{subequations}
\begin{eqnarray}
 t_\text{H}^{(1,\text{F})}\left(x,y;\frac{\muF}{Q^2}\right)
  &=&  t_{\text{H},1}^{(1,\text{F})}\left(x,y\right)
   + t_{\text{H},2}^{(1,\text{F})}\left(x,y;\frac{\muF}{Q^2}\right)\,;
\label{eq:TH1F}\\[1.5mm]
 t_{\text{H},2}^{(1,\text{F})}\left(x,y;\frac{\muF}{Q^2}\right)
 &=& t_{\text{H}}^{(0)}\left(x,y\right)
     \left[2\Big(3 + \ln (\x\y) \Big)\ln \frac{Q^2}{\muF}\right]\,.
\label{eq:TH12F}
\end{eqnarray}
\end{subequations}
Explicit expressions for
$t_{\text{H},1}^{(1,\text{F})}\left(x,y\right)$
and for the color non-singlet part,
$t_\text{H}^{(1,\text{G})}\left(x,y\right)$,
cf.\ Eq.\ (\ref{eq:THNLOpff}),
are supplied in Appendix B
(see Eqs.\ (\ref{eq:TH11F}), (\ref{eq:TH1G})).

The scaled hard-scattering amplitude, Eq.\ (\ref{eq:TH}), truncated at
the NLO and evaluated at the renormalization scale
$\muR=\lambda_\text{R} Q^2$, reads
\begin{eqnarray}
 Q^2 T^\text{NLO}_\text{H}\left(x,y,Q^2;\muF,\lambda_\text{R} Q^2
                          \right)
 = \alpha_s\left(\lambda_\text{R}Q^2\right)\, t_\text{H}^{(0)}(x,y)
  + \frac{\alpha_s^2\left(\lambda_\text{R}Q^2\right)}{4\pi}\,
     C_\text{F}\,
      t_{\text{H},2}^{(1,\text{F})}\left(x,y;\frac{\muF}{Q^2}\right)
~~~
\nonumber \\
 +\ \frac{\alpha_s^2\left(\lambda_\text{R}Q^2\right)}{4\pi}
       \left\{b_0\,t_\text{H}^{(1,\beta)}(x,y;\lambda_\text{R})
          + t_\text{H}^{(\text{FG})}(x,y)
       \right\}\,,~~~~~~~~~~~~~~
\label{eq:TH-mod}
\end{eqnarray}
where we have introduced the shorthand notation
\begin{eqnarray}\label{eq:tHFG}
t_\text{H}^{(\text{FG})}(x,y)\equiv
 C_\text{F}\,t_{\text{H},1}^{(1,\text{F})}(x,y)
 + C_\text{G}\,t_\text{H}^{(1,\text{G})}(x,y)\,.
\end{eqnarray}
To calculate the factorizable part of the pion form factor,
one has to convolute
this expression with the pion DA for each hadron
in the initial
and final state.
In leading twist 2, the pion DA at the normalization scale
$\muO \approx 1$~GeV${}^{2}$
is given by
\begin{eqnarray}
 \varphi_\pi(x,\mu_0^2)
  = 6 x (1-x)
     \left[ 1
          + a_2(\muO) \, C_2^{3/2}(2 x -1)
          + a_4(\muO) \, C_4^{3/2}(2 x -1)
          + \ldots
     \right]\, ,
\label{eq:phi024mu0}
\end{eqnarray}
with all nonperturbative information being encapsulated in the
Gegenbauer coefficients $a_n$.
In this analysis we use those coefficients determined before
by Bakulev, Mikhailov, and Stefanis (BMS) in~\cite{BMS01}
 with the aid of QCD sum rules with nonlocal condensates:
\begin{equation}
  a_{2}^{\rm BMS}=0.20, \quad a_{4}^{\rm BMS}=-0.14,
  \quad a_{n}^{\rm BMS}=0 \:, n > 4 \, ,
\label{eq:BMS-DA}
\end{equation}
where the vacuum quark virtuality
$\lambda_{q}^{2}=0.4$~GeV${}^2$
has been used.
This set of values was found \cite{BMS02,BMS03} to be consistent at
the $1\sigma$ level with the high-precision CLEO data \cite{CLEO98}
on the pion-photon transition form factor, with all other model DAs
being outside---at least---the $2\sigma$ error ellipse (see
\cite{BMS04kg} for the latest compilation of models in comparison with
the CLEO and CELLO \cite{CELLO91} data).
Notice that the particular parameterization (shape) of the pion DA
chosen is irrelevant for the considerations to follow.

\section{Analyticity of Partonic Amplitudes Beyond LO}
\label{sec:analyticity}

\subsection{Analytic Running Coupling in QCD}
\label{subsec:analy-coupl}

The main stumbling block in applying fixed-order perturbation theory
at low momenta $Q^2$ is the non-physical Landau singularity of the
running strong coupling at $Q^2=\Lambda^2$, which entails the
appearance of IR renormalons in the perturbative expansion.
To ensure the analyticity of the coupling in the infrared, one can
follow different strategies
\cite{CS93,BBB95,DMW96,SS97,CMNT96,Magn00,BRS00,Gar01} all based on
the basic assumption that the \emph{physical} coupling should stay IR
finite and analytic in the whole momentum range, though its precise
value at $Q^2=0$ is still a matter of debate
\cite{SS97,Nes03,Ale05,AS00,BP04}.
Imposing the analyticity of the coupling in the sense of Shirkov and
Solovtsov \cite{SS97}, we replace the strong running coupling and its
powers by their analytic versions:
\begin{eqnarray}
 \left[\alpha_s^{(n)}\left(Q^2\right)^m\right]^\text{an}
  \equiv {\cal A}_{m}^{(n)}\left(Q^2\right)~~\text{with}~~
 \left[f\left(Q^2\right)\right]^\text{an}
  =
  \frac{1}{\pi}
   \int_0^{\infty}\!
    \frac{\textbf{Im}\,\big[f(-\sigma)\big]}
         {\sigma+Q^2-i\epsilon}\,
     d\sigma\,,
\end{eqnarray}
where the loop order is explicitly indicated by the superscript $n$
in parenthesis and
\begin{eqnarray}
 {\cal A}_{1}^{(1)}\left(Q^2\right)
  = \frac{4\pi}{b_0}
    \left[\frac{1}{\ln(Q^2/\Lambda^2)}
        + \frac{\Lambda^2}{\Lambda^2-Q^2}
    \right]
 \equiv \bar{\alpha}_s\left(Q^2\right)\, ,
\label{eq:SScouplings}
\end{eqnarray}
with the last step connecting to the SS\ notation \cite{SS97},
and $\alpha_s(0)=4\pi/b_0$.
The two-loop running coupling in standard QCD perturbation theory
can be expressed \cite{Mag99} in terms of the Lambert function
$W_{-1}$ to read
\begin{eqnarray}
 \label{eq:alphaexact}
  \as^{(2)}\left(Q^2\right)
  = -\frac{4\pi}{b_0 c_1}
      \left[1
          + W_{-1}\left(-\frac{1}{c_1e}
                      \left(\frac{\Lambda^2}{Q^2}
                      \right)^{1/c_1}
             \right)
      \right]^{-1}\,.
\end{eqnarray}
For some more explanations we refer the interested reader to
\cite{BMS02}, Appendix C, Eqs.\ (C15) and (C20).
Then, the analytic image of the $k$th power of the coupling
\cite{DVS00unp} is obtained from the dispersion relation
\begin{eqnarray}
 \acal_{k}^{(2)}\left(Q^2\right)
  = \frac{1}{\pi}\int_0^{\infty}
     d\sigma\frac{\rho_k^{(2)}(\sigma)}{\sigma+Q^2-i\epsilon}\,
\label{eq:2loop-A_k}
\end{eqnarray}
with the spectral density
\begin{eqnarray}
\label{eq:ro_k}
  \rho_k^{(2)}(t)
   = \left(\frac{4\pi}{b_0 c_1}\right)^{k}
     \Im\left(-\frac{1}{1+W_{1}(z(t))}\right)^{k}\,.
\end{eqnarray}
In the numerical calculations below, we use an approximate form
suggested in \cite{BPSS04}:
\begin{eqnarray}
\label{eq:asb_2-Appro}
  \acal_{1}^{(2,\text{fit})}\left(Q^2\right)
  &=& \frac{4\pi}{b_0}\,
  \left\{\frac{1}{\ell
  \left[\ln\left(Q^2/\Lambda_{21}^2\right),\,c_{21}^\text{fit}\right]}
  + \frac{1}{1 - \exp(\ell
  \left[\ln\left(Q^2/\Lambda_{21}^2\right),\,c_{21}^\text{fit}\right])}
      \right\}\,;\\ 
\label{eq:A2_2-Appro}
  \acal_{2}^{(2,\text{fit})}\left(Q^2\right)
  &=& \left(\frac{4\pi}{b_0}\right)^2
  \left\{\frac{1}{\ell
  \left[\ln\left(Q^2/\Lambda_{22}^2\right),c_{22}^\text{fit}\right]^2}
  - \frac{\exp(\ell
  \left[\ln\left(Q^2/\Lambda_{22}^2\right),c_{22}^\text{fit}\right])}
  {\left[1 - \exp(\ell
  \left[\ln\left(Q^2/\Lambda_{22}^2\right),c_{22}^\text{fit}\right])
  \right]^2}
      \right\}\,,~~~
\end{eqnarray}
where the values of the fit parameters are listed in Table
\ref{tab:tab-2} and
\begin{eqnarray}
 \ell[L,c] \equiv L + c\,\ln\sqrt{L^{2}+4\pi^2}\,.
\end{eqnarray}
\begin{table}[t]
\caption{\label{tab:tab-2}
 Parameters entering Eqs.\ (\protect{\ref{eq:asb_2-Appro}})
 and (\protect{\ref{eq:A2_2-Appro}})
 for the value $\Lambda_\text{QCD}^{N_f=3}=400$~MeV.}
\begin{ruledtabular}
\begin{tabular}{ccccc}
Parameters & $c_{21}^\text{fit}$ & $\Lambda_{21}$ & $c_{22}^\text{fit}$
                                                  &  $\Lambda_{22}$ \\
                                                  \hline
Values     & $-1.015$            & $67$~MeV       & $-1.544$   &
                                                    $34.5$~MeV \\
\end{tabular}
\end{ruledtabular}
\end{table}

\subsection{``Analytization'' Procedures}
\label{subsec:analy-proc}

Let us now see how analyticity can be implemented on the parton-level
pion form factor in NLO accuracy of perturbative QCD.
We discuss three ``analytization'' procedures:\footnote{%
One should not worry about the factor $1/Q^2$ because under
``analytization'' it reproduces itself, i.e.,
$\left[\left[f\left(Q^2\right)\right]^\text{an}/Q^2\right]^\text{an}
 =\left[f\left(Q^2\right)\right]^\text{an}/Q^2$.}
\begin{itemize}
\item \emph{Naive} ``analytization'' \cite{SSK99,SSK00,BPSS04}
\begin{eqnarray}
 \left[Q^2 T_\text{H}\left(x,y,Q^2;\muF,\lambda_\text{R} Q^2\right)
 \right]_\text{SS}^\text{naive-an}
  = {\cal A}_{1}^{(2)}(\lambda_\text{R} Q^2)\,  t_\text{H}^{(0)}(x,y)
~~~~~~~~~~~~~~~~~~~~~~~~~~~~~
 \nonumber\\ [0.5cm]
 + \frac{\left({\cal A}_{1}^{(2)}(\lambda_\text{R} Q^2)
         \right)^2}{4\pi}\,
       \left[b_0\,t_\text{H}^{(1,\beta)}(x,y;\lambda_\text{R})
           + t_\text{H}^{(\text{FG})}(x,y)
           + C_\text{F}\,
             t_{\text{H},2}^{(1,\text{F})}
             \left(x,y;\frac{\muF}{Q^2}\right)
       \right]\,.~~~~~
\end{eqnarray}
\item \emph{Maximal} ``analytization'' \cite{BPSS04}
\begin{eqnarray}
 \left[Q^2 T_\text{H}\left(x,y,Q^2;\muF,\lambda_\text{R} Q^2\right)
 \right]_\text{SS}^\text{max-an}
 = {\cal A}_{1}^{(2)}(\lambda_\text{R} Q^2)\,  t_\text{H}^{(0)}(x,y)
~~~~~~~~~~~~~~~~~~~~~~~~~~~~~~~~~~~
 \nonumber\\ [0.5cm]
 + \frac{\acal_{2}^{(2)}(\lambda_\text{R} Q^2)}{4\pi}\,
       \left[b_0\,t_\text{H}^{(1,\beta)}(x,y;\lambda_\text{R})
           + t_\text{H}^{(\text{FG})}(x,y)
           + C_\text{F}\,
             t_{\text{H},2}^{(1,\text{F})}
             \left(x,y;\frac{\muF}{Q^2}\right)
       \right]\,.~~~~~
\label{eq:TH-SS-max}
\end{eqnarray}
\item \emph{Amplitude} ``analytization'' proposed by Karanikas
and Stefanis in \cite{KS01,Ste02}.
\end{itemize}
The first method replaces $\alpha_s$ and its powers by the
Shirkov--Solovtsov analytic coupling \cite{SS97} and its powers,
whereas the second one uses for the powers of $\alpha_s$ their own
analytic images, transforming this way the power-series expansion in
$[\alpha_s\left(Q^2\right)]^n$ in a functional expansion in terms of
the functions
${\cal A}_{n}\left(Q^2\right)$ \cite{SS99,Shi00}.
Imposing analyticity in the sense of Karanikas--Stefanis \cite{KS01},
differs from the previous two approaches in that it demands the
\emph{whole} partonic amplitude has the correct analytical behavior
as a function of $Q^2$.
This entails the ``analytization'' of terms of the form
$\left(\alpha_{s}^{(n)}\left(Q^2\right)
 \right)^m\ln (Q^2/\mu_{\rm F}^2)$,
which appear in exclusive amplitudes at NLO of QCD perturbation
theory and contain an additional scale, $\mu_{\rm F}^2$.
There are, in principle, two possibilities how to proceed any further.
One option is provided by setting $\muF\simeq Q^2$ and then face the
problem of ``analytization'' of terms like
$\left[\alpha_{s}\left(Q^2\right)/\alpha_{s}(\muO)
\right]^{\eta}$,
where
$\eta=\gamma_n^{(0)}/(2b_0)$ is a fractional number, as discussed in
\cite{BMS05}.
Another possibility is to fix the factorization scale $\muF$ at some
value and then to redefine the original Shirkov--Solovtsov
``analytization'' procedure in order to take the dispersive image of
the coupling (or of its powers) together with these logarithmic terms.
This second route is followed in the present work.
It is important to note that the KS ``analytization'' procedure reduces
in LO of fixed-order perturbation theory to the \emph{maximal} one, as
shown in \cite{KS01}, provided evolution effects of the pion
distribution amplitudes are ignored.

\begin{figure}[t]
 \centerline{\includegraphics[width=\textwidth]{
  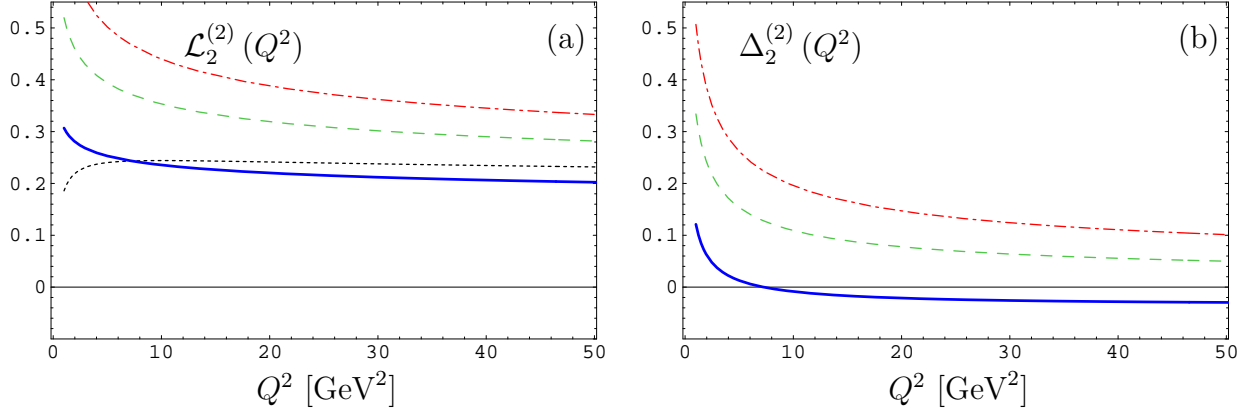}}%
  \caption{\label{fig:comp_Max_KS}\footnotesize
    (a): Results for the analyticized logarithmic term,
    ${\cal L}_{2}^{(2)}\left(Q^2\right)$,
    using three different ``analytization'' procedures:
    the one-loop approximate KS ``analytization'',
    ${\cal L}_{2}^{(1)\,\text{approx}}\left(Q^2\right)$
    (red dash-dotted line),
    the two-loop approximate KS ``analytization'',
    ${\cal L}_{2}^{(2)\,\text{approx}}\left(Q^2\right)$
    (green dashed line),
    and the exact two-loop BMKS\ ``analytization''
    ${\cal L}_{2}^{(2)}\left(Q^2\right)$ (blue solid line).
    For comparison, we also show here the corresponding
    \emph{maximal} ``analytization'' curve (dotted line).
    (b): Results are shown for the corresponding analyticized
    contributions  $\Delta_{2}^{(2)}\left(Q^2\right)$,
    using the same three ``analytization'' procedures for
    ${\cal L}_{2}^{(2)}\left(Q^2\right)$
    as in panel (a).}
\end{figure}

Applying now this generalized ``analytization'' concept, we get
\begin{eqnarray}
 \left[Q^2 T_\text{H}\left(x,y,Q^2;\muF,\lambda_\text{R} Q^2\right)
  \right]_\text{KS}^\text{an}
 = {\cal A}_{1}^{(2)}(\lambda_\text{R} Q^2)\,  t_\text{H}^{(0)}(x,y)
   ~~~~~~~~~~~~~~~~~~~~~~~~~~~~~~~~~~~~~~~~~~~~~\nonumber\\ 
 +\ \frac{{\cal A}_{2}^{(2)}(\lambda_\text{R} Q^2)}{4\pi}\,
       \left(b_0\,t_\text{H}^{(1,\beta)}(x,y;\lambda_\text{R})
           + t_\text{H}^{(\text{FG})}(x,y)
       \right)~~~~~~~~~~~~~\nonumber\\
 + \left[ \frac{\left(\alpha_{s}^{(2)}(\lambda_\text{R} Q^2)
                  \right)^2}{4\pi}\,
             C_\text{F}\, t_\text{H}^{(0)}(x,y)
             \left(6 + 2\ln (\x\y)\right)\ln \frac{Q^2}{\muF}
     \right]_\text{KS}^\text{an}.~~~~
\label{eq:TH-KS-2}
\end{eqnarray}
In order to have the same scale argument in the logarithmic term as
in the running coupling, we substitute
$\ln(Q^2/\muF)=
 \ln (\lambda_\text{R} Q^2/\Lambda^2) -
 \ln (\lambda_\text{R}\muF/\Lambda^2)$
to obtain
\begin{eqnarray}
 \left[Q^2 T_\text{H}\left(x,y,Q^2;\muF,\lambda_\text{R} Q^2\right)
 \right]_\text{KS}^\text{an}
  = {\cal A}_{1}^{(2)}(\lambda_\text{R} Q^2)\,  t_\text{H}^{(0)}(x,y)
  ~~~~~~~~~~~~~~~~~~~~~~~~~~~~~~~~~~~~~~~~~~~~
  \nonumber\\ 
  ~~~~~+\ \frac{{\cal A}_{2}^{(2)}(\lambda_\text{R} Q^2)}{4\pi}\,
       \left[b_0\,t_\text{H}^{(1,\beta)}(x,y;\lambda_\text{R})
           + t_\text{H}^{(\text{FG})}(x,y)
           - C_\text{F}\, t_\text{H}^{(0)}(x,y)
              \left(6 + 2\ln (\x\y)\right)
               \ln \frac{\lambda_\text{R}\muF}{\Lambda^2}
       \right]
 \nonumber\\ 
 +\ \left[\frac{\left(\alpha_{s}^{(2)}(\lambda_\text{R} Q^2)
                  \right)^2}{4\pi}\,
             C_\text{F}\, t_\text{H}^{(0)}(x,y)
              \left(6 + 2\ln (\x\y)\right)
               \ln \frac{\lambda_\text{R} Q^2}{\Lambda^2}
     \right]_\text{KS}^\text{an}.~~~~~~~~~~~~~~~~~~~~~~~~~~~~~~~~\
 \label{eq:TH-KS-3}
\end{eqnarray}
Finally, we arrive at
\begin{eqnarray}
 \left[Q^2 T_\text{H}(x,y,Q^2;\muF,\lambda_\text{R} Q^2)
 \right]_\text{KS}^\text{an}
  = {\cal A}_{1}^{(2)}(\lambda_\text{R} Q^2)\, t_\text{H}^{(0)}(x,y)
 ~~~~~~~~~~~~~~~~~~~~~~~~~~~~~~~~~~~~~~~~~~~~~~
 \nonumber\\ 
  ~~~~+\ \frac{\acal_{2}^{(2)}(\lambda_\text{R} Q^2)}{4\pi}\,
       \left[b_0\,t_\text{H}^{(1,\beta)}(x,y;\lambda_\text{R})
           + t_\text{H}^{(\text{FG})}(x,y)
           + C_\text{F}\,
             t_{\text{H},2}^{(1,\text{F})}
             \left(x,y;\frac{\muF}{Q^2}\right)
       \right]
       \nonumber\\
  +\ \frac{\Delta_{2}^{(2)}
  \left(\lambda_\text{R} Q^2\right)}{4\pi}\,
      \left[C_\text{F}\, t_\text{H}^{(0)}(x,y)  \,
             \left(6 + 2 \ln(\bar{x}\bar{y})\right)
      \right]\,,~~~~~~~~~~~~~~~~~~~~~~~~~~~~~~~
 \label{eq:TH-KS-6}
\end{eqnarray}
where the deviation from Eq.\ (\ref{eq:TH-SS-max}) is encoded in
the term
\begin{eqnarray}\label{eq:delta2-2}
 \Delta_{2}^{(2)}\left(Q^2\right)
  &\equiv&
   {\cal L}_{2}^{(2)}\left(Q^2\right)
    - {\cal A}_{2}^{(2)}\left(Q^2\right)\,\ln\left[Q^2/\Lambda^2\right]
\end{eqnarray}
with
\begin{eqnarray}\label{eq:Log_Alpha_2_KS}
 {\cal L}_{2}^{(2)}\left(Q^2\right)
  &\equiv&
   \left[\left(\alpha_{s}^{(2)}\left(Q^2\right)\right)^2
         \ln\left(\frac{Q^2}{\Lambda^2}\right)
   \right]_\text{KS}^\text{an}
  =  \frac{4\pi}{b_0}\,
     \left[\frac{\left(\alpha_{s}^{(2)}\left(Q^2\right)\right)^2}
                 {\alpha_{s}^{(1)}\left(Q^2\right)}
      \right]_\text{KS}^\text{an}\, .
\end{eqnarray}
It is important to distinguish between the two contributions in
Eq.\ (\ref{eq:delta2-2}).
The first amounts to the ``analytization'' of the product of the
coupling with a logarithm, or equivalently of fractional powers of
the coupling, as shown in \cite{BMS05}.
The second bears an additional logarithmic dependence on the momentum
scale $Q^2$ relative to the expression obtained with the \emph{maximal}
``analytization'' procedure.
The subscript KS in the last equation signifies that this expression
should be analyticized according to the KS prescription.
To obtain a clearer idea of its meaning and demonstrate its essence,
the ``analytization'' is performed in three incremental steps.
First, a simplified version of this expression is considered, which
results by provisionally replacing the two-loop coupling in the
numerator by its one-loop counterpart.
Then, the ratio of the couplings after ``analytization'' reduces to
(dash-dotted line in Fig.\ \ref{fig:comp_Max_KS}a)
\begin{eqnarray}\label{eq:Log_Alpha_2_KS_Approx1}
  {\cal L}_{2}^{(1)\,\text{approx}}\left(Q^2\right)
   = \frac{4\pi}{b_0}\,
      {\cal A}_{1}^{(1)}\left(Q^2\right)\,.
\end{eqnarray}

Second, we discuss an analogous situation, in which the one-loop
coupling in the denominator is (inconsistently) traded for its
two-loop counterpart.
In this case, the ratio of the couplings after ``analytization''
becomes (dashed line in Fig.\ \ref{fig:comp_Max_KS}a)
\begin{eqnarray}\label{eq:Log_Alpha_2_KS_Approx2}
  {\cal L}_{2}^{(2)\,\text{approx}}\left(Q^2\right)
   = \frac{4\pi}{b_0}\,
      {\cal A}_{1}^{(2)}\left(Q^2\right)\,.
\end{eqnarray}
Finally, we provide the exact result for the KS ``analytization'' of
expression (\ref{eq:Log_Alpha_2_KS}) (solid line in Fig.\
\ref{fig:comp_Max_KS}a), with the derivation presented in Appendix
\ref{app:HSAnlo}, while more general expressions are given in
\cite{BMS05}:
\begin{eqnarray}\label{eq:Log_Alpha_2_BMKS}
  {\cal L}_{2}^{(2)}\left(Q^2\right)
   = \frac{4\pi}{b_0}\,
      \left[{\cal A}_{1}^{(2)}\left(Q^2\right)
      + c_1\,\frac{4\pi}{b_0}\,f_{\cal L}\left(Q^2\right)
      \right]\, ,
\end{eqnarray}
where
\begin{eqnarray}\label{eq:f_MS}
  f_{\cal L}\left(Q^2\right)
   = \sum_{n\geq0}
      \left[\psi(2)\zeta(-n-1)-\frac{d\zeta(-n-1)}{dn}\right]\,
       \frac{\left[-\ln\left(Q^2/\Lambda^2\right)
             \right]^n}{\Gamma(n+1)}
\end{eqnarray}
and $\zeta(z)$ is the Riemann zeta-function.
Equation (\ref{eq:delta2-2}) is illustrated in Fig.\
\ref{fig:comp_Max_KS}b for the different expressions of
${\cal L}_{2}\left(Q^2\right)$
given by Eqs.\ (\ref{eq:Log_Alpha_2_KS_Approx1}),
(\ref{eq:Log_Alpha_2_KS_Approx2}), and (\ref{eq:Log_Alpha_2_BMKS}),
using the same line designations as in Fig.\ \ref{fig:comp_Max_KS}a.
Let us close this discussion by commenting that in the region where
there are experimental data available \cite{FFPI73,JLAB00}
(i.e., well below 10 GeV${}^{2}$),
Eq.\ (\ref{eq:delta2-2}) is governed by
${\cal L}_{2}^{(2)}\left(Q^2\right)$, which entails a small enhancement
of the hard-scattering amplitude for $Q^2\leq 7.25$~GeV${}^{2}$.

\section{Factorized Pion Form Factor at
         NLO---Standard and Analyticized}
\label{sec:FFanalytic}

The calculation of the factorized pion form factor proceeds in terms
of Eq.\ (\ref{eq:pff-Fact}) and involves the convolution of expression
(\ref{eq:TH-SS-max}) for the \emph{maximal} ``analytization'' case,
or expression (\ref{eq:TH-KS-6}) for the KS ``analytization'' case
with the pion DA for which we employ in both cases the BMS\
parameterization \cite{BMS01}, as discussed in Sec.\
\ref{sec:standardQCD}.
On that basis, we can obtain the scaled, factorized part of the pion
form factor,
$Q^2F_{\pi}^\text{Fact}\left(Q^2;\muR=\lambda_\text{R}Q^2\right)$,
using Eq.\ (\ref{eq:TH-mod}) and the following set of
substitutions:\footnote{%
Here, we write for the sake of brevity
$a_2=a^\text{BMS}_2(\muF)$ and $a_4=a^\text{BMS}_4(\muF)$
and use the values given in Eq.\ (\ref{eq:BMS-DA}).}
\begin{eqnarray}
 t_\text{H}^{(0)}(x,y) &\rightarrow&
  8\,\pi\,f_{\pi}^2\,
     \left(1 + a_2 + a_4\right)^2\,;
\label{eq:tHLO}\\
 -\,t_\text{H}^{(0)}(x,y)\,\ln \x\y &\rightarrow&
  8\,\pi\,f_{\pi}^2\,
     \left(1 + a_2+ a_4\right)\,
       \left[3 + (43/6) a_2 + (136/15) a_4\right]\,;~~~
\label{eq:tHLOLog}\\
 t_\text{H}^{(\text{FG})}(x,y)
          &\rightarrow&
  8\,\pi\,f_{\pi}^2\,
    \big[- 15.67 - a_2 \left(21.52 - 6.22\, a_2\right)\nonumber\\
  & &  ~~~~~~~~~ - a_4 \left(7.37 - 37.40\, a_2 - 33.61\, a_4 \right)
      \big]\,.
\label{eq:Q2pff1FG}
\end{eqnarray}

Notice that evolving the BMS\ pion DA from the initial scale $\muO$
to the scale $\muF$ at the NLO level will generate higher Gegenbauer
harmonics of the form
$x\, \x C^{3/2}_{2n}(2x-1)$ with $n\geq3$.
However, we have shown in \cite{BPSS04} (see also \cite{BS05}) that
for the calculation of the pion form factor it is actually sufficient
to restrict ourselves to the LO evolution and neglect NLO evolution
effects.
Hence, for our purposes in the present analysis, we set
\begin{eqnarray}
 a_{2n}(\muF)
  = a_{2n}(\muO)
    \left[\frac{\alpha_{s}(\muF)}{\alpha_{s}(\muO)}
    \right]^{\gamma_n^{(0)}/(2b_0)}\, .
\label{eq:LOevo}
\end{eqnarray}
The lowest-order anomalous dimensions can be represented in closed
form by
\begin{eqnarray}
 \gamma_n^{(0)}
   = 2 C_\text{F}\left[4 S_1(n+1)
                 - 3 - \frac{2}{(n+1) (n+2)}
           \right]
\label{eq:gamma0}
\end{eqnarray}
with $S_1(n+1)=\sum_{i=1}^{n+1} 1/i =\psi(n+2)-\psi(1)$,
while the function $\psi(z)$
is defined as $\psi(z)= d\ln\Gamma(z)/d z$.

Following the master plan for ``analytization'', exposed in the
previous section, we obtain the following expressions for the
factorized pion form factor:
\begin{itemize}
\item\emph{Naive} ``analytization'' \cite{BPSS04,SSK99,SSK00}:
\begin{eqnarray}
 \left[ F_{\pi}^\text{Fact}(Q^2; \lambda_{\rm R}Q^2)
 \right]_\text{NaivAn}
  &=&  {\cal A}_{1}^{(2)}(\lambda_{\rm R}Q^2)\,
      {\cal F}_{\pi}^\text{LO}\left(Q^2\right)\nonumber\\
  &+&  \frac{1}{\pi}\,
        \left[{\cal A}_{1}^{(2)}(\lambda_{\rm R}Q^2)\right]^2\,
        {\cal F}_{\pi}^{\text{NLO}}\left(Q^2,\muF;\lambda_{\rm R}
                                   \right)\,.
\label{eq:pffNaivAn}
\end{eqnarray}
\item \emph{Maximal} ``analytization'' \cite{BPSS04}:
\begin{eqnarray}
 \left[ F_{\pi}^\text{Fact}(Q^2; \lambda_{\rm R}Q^2)
 \right]_\text{MaxAn}
 &=& {\cal A}_{1}^{(2)}(\lambda_{\rm R}Q^2)\,
           {\cal F}_{\pi}^\text{LO}\left(Q^2\right)
    \nonumber \\
 &+& \frac{1}{\pi}\,
      {\cal A}_{2}^{(2)}(\lambda_{\rm R}Q^2)\,
      {\cal F}_{\pi}^\text{NLO}\left(Q^2,\muF;\lambda_{\rm R}\right)\,.
\label{eq:pffMaxAn}
\end{eqnarray}
\item KS \emph{amplitude} ``analytization'' (this work)---cf.\ Eqs.\
(\ref{eq:TH-KS-6}) and (\ref{eq:Log_Alpha_2_KS}):
\begin{eqnarray}
 \left[ F_{\pi}^\text{Fact}(Q^2; \lambda_{\rm R}Q^2)
 \right]_\text{KS}
 &=& {\cal A}_{1}^{(2)}(\lambda_{\rm R}Q^2)\,
           {\cal F}_{\pi}^\text{LO}\left(Q^2\right)
   \nonumber \\
 &+& \frac{1}{\pi}\,
      {\cal A}_{2}^{(2)}(\lambda_{\rm R}Q^2)\,
       {\cal F}_{\pi}^\text{NLO}\left(Q^2,\muF;\lambda_{\rm R}\right)
\nonumber\\
 &+& \frac{\Delta_{2}^{(2)}
           \left(\lambda_\text{R} Q^2\right)}
          {\pi}\,
    \Delta_\text{F}{\cal F}_{\pi}^{\text{NLO}}\left(Q^2\right)\,.~~~~~~
\label{eq:pffKSAn}
\end{eqnarray}
\end{itemize}

\begin{figure}[t]
 \centerline{\includegraphics[width=\textwidth]{
  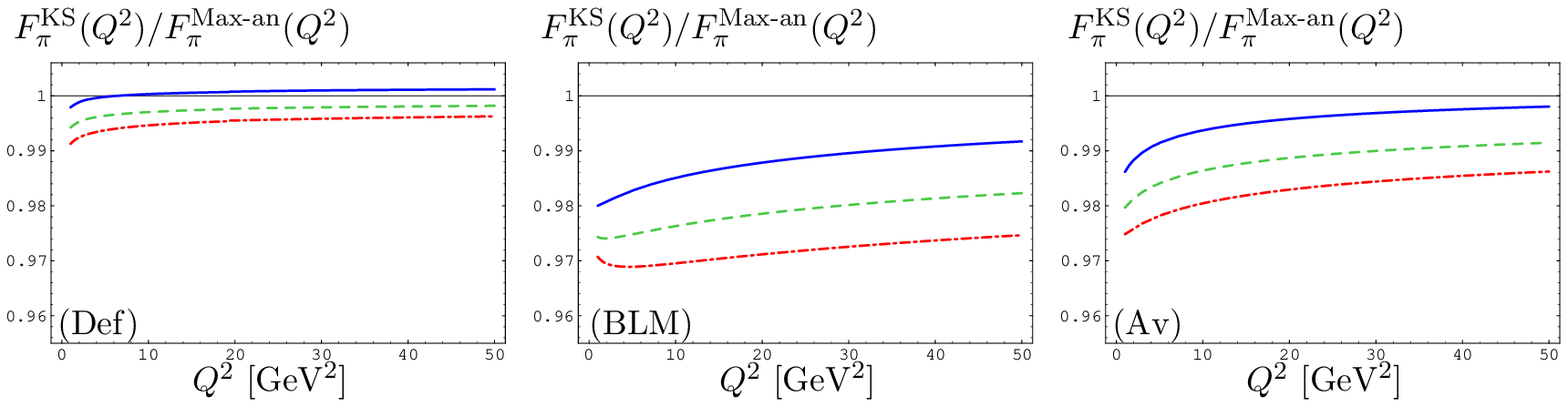}}%
  \caption{\label{fig:ratio_Max_KS}\footnotesize
    Results for the ratio of the factorized pion form factors, using
    two different ``analytization'' procedures:
    KS ``analytization'' and ``maximal analytization'',
    $F^\text{KS}_{\pi}\left(Q^2
                      \right)/F^\text{Max-an}_{\pi}\left(Q^2\right)$.
    The designations are:
    red dash-dotted line---one-loop approximation
    of the KS logarithmic term
    (${\cal L}_{2}^{(1)\,\text{approx}}\left(Q^2\right)$);
    green dashed line---two-loop approximation
    (${\cal L}_{2}^{(2)\,\text{approx}}\left(Q^2\right)$);
    blue solid line---exact two-loop KS ``analytization''
    (${\cal L}_{2}^{(2)}\left(Q^2\right)$).
    Left panel: default scale setting
    ($\lambda=1$); middle panel: BLM\ scale setting; right panel:
    $\alpha_V$ scheme.
    The factorization scale $\mu_\text{F}^2$ is set equal to
    $5.76$~GeV$^2$ \cite{SY99}.}
\end{figure}

Here we use the following notations:
\begin{eqnarray}
 {\cal F}_{\pi}^\text{LO}\left(Q^2\right)
  &=&
  \frac{8\,\pi\,f_{\pi}^2}{Q^2}\,
     \left(1 + a_2 + a_4\right)^2\,;\\
 ~~~{\cal F}_{\pi}^{\text{NLO}}\left(Q^2,\muF;\lambda_{\rm R}\right)
  &=&
  \frac{2\,\pi\,f_{\pi}^2}{Q^2}\,
  \Big[b_0\left(1 + a_2 + a_4\right)^2
          \left(\ln \lambda_\text{R}
               - \ln \lambda_{\text{BLM}}(a_2,a_4)
          \right)
          - 15.67
\nonumber\\
 &&~~~~~~ - a_2 \left(21.52 - 6.22\, a_2\right)
            - a_4 \left(7.37 - 37.40\, a_2 - 33.61\, a_4 \right)
 \Big]\nonumber\\
 &+& \Delta_\text{F}{\cal F}_{\pi}^{\text{NLO}}\left(Q^2\right)
      \ln\frac{Q^2}{\muF}
\end{eqnarray}
and we explicitly display the contribution due to
$t_{\text{H},2}^{(1,\text{F})}\left(x,y;\muF/Q^2\right)$,
see Eq.\ (\ref{eq:TH12F}):
\begin{eqnarray}
 \Delta_\text{F}{\cal F}_{\pi}^{\text{NLO}}(Q^2)
  &=& - \frac{2\,\pi\,f_{\pi}^2}{Q^2}\,
         C_\text{F}\,
          \left(1 + a_2 + a_4\right)
           \left[ (25/3) a_2 + (182/15) a_4 \right]\,.
 \end{eqnarray}
In order to make our formulas more compact, we implement the BLM scale:
\begin{eqnarray}
  \lambda_{\text{BLM}}(a_2,a_4)
   = \exp \left[- \frac{5}{3}
                - \frac{3 +\displaystyle (43/6) a_2 + (136/15) a_4}
                       {1 + a_2 + a_4}
          \right]\,.
\label{eq:aBLMpff}
\end{eqnarray}
\begin{figure}[b]
 \centerline{\includegraphics[width=\textwidth]{
  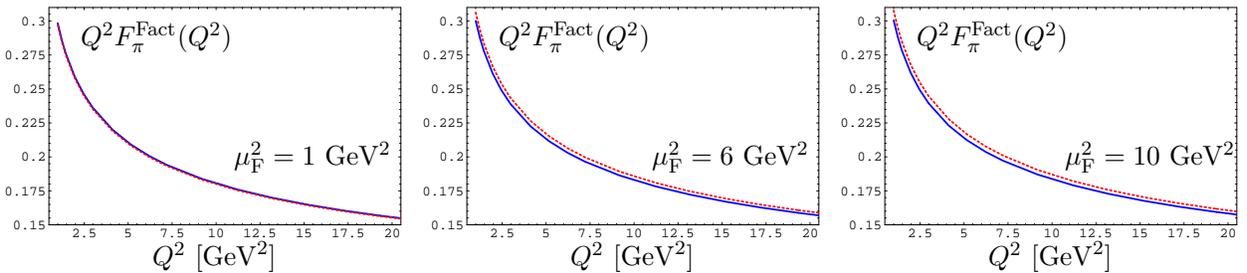}}%
  \caption{\label{fig:FF_Fact_Max_KS_MuF}\footnotesize
    Results for the factorized pion form factors, using two different
    ``analytization'' procedures:
    KS ``analytization'' (blue solid line) and ``maximal
    analytization'' (dotted red line) for different values of the
    factorization scale:
    in the left panel $\muF=1$~GeV$^2$, in the middle
    one--$\muF=1$~GeV$^2$, and in the right one--$\muF=10$~GeV$^2$.
    For all panels we show results for the  BLM\ scale setting.}
\end{figure}

\begin{figure}[h]
 \centerline{\includegraphics[width=0.5\textwidth]{
  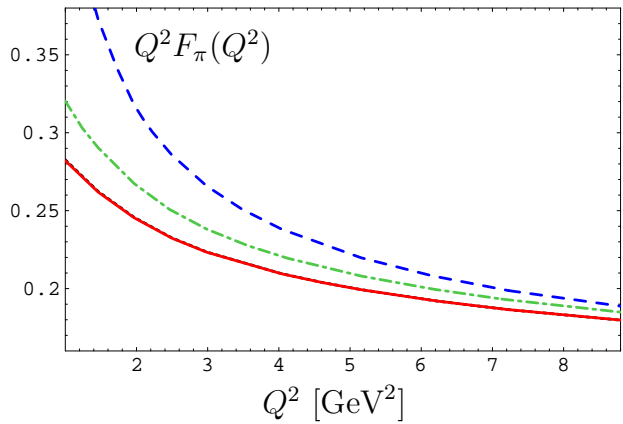}}%
  \caption{\label{fig:pion-form-factor}\footnotesize
    Results for the factorized pion form factor, scaled with $Q^2$,
    and assuming the default scale setting ($\muR=Q^2$) in standard
    perturbation theory and APT.
    The latter is implemented in terms of two different
    ``analytization'' procedures:
    \emph{naive} ``analytization'' and \emph{maximal}
    ``analytization''.
    The designations are:
    blue dashed line---standard perturbation theory;
    green dash-dotted line---\emph{naive} APT;
    red solid line---\emph{maximal} APT.
    The prediction obtained with the KS ``analytization'' is too close
    to that found with the \emph{maximal} one to differentiate these
    curves graphically.
    The factorization scale $\mu_\text{F}^2$ is set equal to
    $5.76$~GeV$^2$.}
\end{figure}

The ``analytization'' augmented perturbation theory works very well.
This is illustrated by the results in Figs.\ \ref{fig:ratio_Max_KS},
\ref{fig:FF_Fact_Max_KS_MuF}, and \ref{fig:pion-form-factor}.
The first of these figures compares the specific issues of the KS
``analytization'' procedure relative to those of the \emph{maximal}
one for the ratio of the corresponding factorized form factors.
A few words are in order here.
One sees that using the default \MS scheme, the KS ``analytization''
procedure yields a result almost coincident with that provided by the
\emph{maximal} one.
On the other hand, in the BLM\ scheme and also in the $\alpha_V$
scheme, the KS prediction is smaller by a few percent.
Moreover, one observes by comparison with Fig.\ 11, right panel in
Ref.\ \cite{BPSS04} that the BLM\ prediction, which in the
\emph{maximal} procedure was the largest one, becomes in the case of
the KS prescription comparable with the prediction of the default
scheme.
As a result, the inherent theoretical uncertainties due to the
involved perturbative parameters, defining a renormalization scheme and
scale setting, are further reduced.
A second important feature of the KS procedure is that the
dependence of $F_{\pi}^{\rm Fact}(Q^2)$ on the factorization
scale is almost diminished, as indicated in Fig.\
\ref{fig:FF_Fact_Max_KS_MuF}.
Indeed, varying the factorization scale from 1~GeV${}^2$ to
10~GeV${}^2$, the form factor changes by a mere 1.5 percent.
Even setting the factorization scale to the theoretical value of
50~GeV${}^{2}$, the induced variation in the form-factor magnitude
reaches just the level of about 2.5 percent.
In the case of the \emph{maximal} ``analytization'' procedure, the
dependence on the factorization scale is also a mild one, but the
corresponding variation is, in round terms, two times larger.

The fourth figure demonstrates the impact of ``analytization'' on the
factorized pion's electromagnetic form factor, using various
``analytization'' prescriptions.
The dashed line denotes the prediction obtained with standard QCD
perturbation theory in the \MS scheme and applying the default scale
setting $\mu_{\rm R}^{2}=Q^{2}$.
The \emph{naive} ``analytization'' prediction is represented by the
dash-dotted line and the analogous one for the \emph{maximal}
``analytization'' by the solid line below it.
The result of the calculation according to the KS ``analytization''
practically coincides with that of the \emph{maximal} one.
This behavior is also reflected in Fig.\ \ref{fig:ratio_Max_KS},
where we see that the differences among the three ``analytization''
procedures are of the order of a few percent in the whole $Q^2$ range
considered.

Note that as regards the whole pion form factor, i.e., taking into
account also the soft part, the differences would be further reduced.
For full details the reader is referred to \cite{BPSS04}.

\section{Summary and Conclusions}
\label{sec:concl}

We have discussed different ``analytization'' procedures to ensure
the analyticity of the factorized electromagnetic pion form factor
at NLO of QCD perturbation theory.
The main features and relative merits of each ``analytization''
concept following from the presented analysis are:
\begin{itemize}
\item
The \emph{naive} ``analytization \cite{SSK99,SSK00} retains the
power-series expansion of perturbative QCD, but replaces
$(\alpha_s^{(n)}(Q^{2}))^{m}$ by
$({\cal A}_{1}^{(n)}(Q^{2}))^{m}$.
As it was shown in \cite{SSK99,SSK00}, this reduces the value of the
NLO correction, though the sensitivity to the renormalization scheme
adopted and the renormalization scale-setting chosen is still
substantial, resulting into a rather strong variation of the
form-factor predictions \cite{BPSS04}.
Moreover, this procedure does not respect nonlinear relations of the
coupling because these correspond to different dispersive
images.

\item
The \emph{maximal} ``analytization'' \cite{BPSS04} trades the
power-series expansion for a functional non-power-series expansion in
terms of ${\cal A}_{m}^{(n)}(Q^{2})$ \cite{SS97,Shi00,SS01},
minimizing the variation of the form-factor predictions owing to the
renormalization scheme and scale setting.
It is, however, insufficient to cure logarithms of the momentum
scale multiplying the running coupling.
Such terms modify the spectral density, i.e., the discontinuity across
the cut along the negative real axis and have therefore to be taken
into account.

\item
Applying the ``analytization'' procedure at the level of the
partonic amplitude itself \cite{KS01,Ste02}, bears all advantages of
the \emph{maximal} ``analytization'' plus a further reduced dependence
on the perturbative scales---especially the dependence on the
factorization scale.
This has been verified by explicit calculation.
We have employed the \MS scheme with various scale settings and also
the $\alpha_V$ scheme.
In addition, we have varied the factorization scale in the range
$1-10$~GeV${}^2$.
While the predictions for the factorized pion form factor, calculated
with the \emph{maximal} procedure, were affected by this variation on
the level of 3\%, their counterparts, derived with the KS prescription,
were influenced by less than 1\%.
Though the KS method does not really ``gain up'' relative to the
\emph{maximal} ``analytization'' procedure with respect to the
factorized pion form factor, as one observes from Fig.\
\ref{fig:pion-form-factor}, it is able to further improve the
perturbative treatment because it extends the notion of analyticity
to non-integer powers of the strong running coupling---FAPT.
Such powers become relevant when one has to calculate the analytic
image of powers of the strong coupling in combination with logarithms,
the latter first appearing at NLO of fixed-order perturbation theory,
or in terms of evolution factors \cite{BMS05}.
Hence, the KS ``analytization'' requirement treats all logarithms
that have a non-zero spectral density, and hence modify the discontinuity
across the cut along the negative real axis, on the same footing and
irrespective of their source being it the running coupling (and its powers),
or logarithms entailed by ERBL\ or DGLAP\ evolution.
\end{itemize}

In conclusion, the KS ``analytization'' enables the variation of the
factorization scale and the choice of various renormalization schemes
and scale settings, including the BLM\ one, with undiminished quality
of the theoretical predictions from scheme (scale) to scheme (scale),
virtually eliminating the dependence on such parameters and upgrading
the \MS scheme to an optimized factorization and renormalization
scheme.
From a broader perspective one may interpret these findings as
indicating that the analyticity of the partonic three-point function
is as important and fundamental as the underlying symmetries of the
theory and should be preserved together with them in the maximal
possible way.

\acknowledgments
We wish to thank Sergey Mikhailov for valuable discussions and
comments.
Two of us (A.P.B.\ and A.I.K.) are indebted to Prof.\ Klaus Goeke
for the warm hospitality at Bochum University, where the major part
of this investigation was carried out.
This work was supported in part by the Deutsche Forschungsgemeinschaft,
the Verbundforschung des Bundesministeriums f\"{u}r Bildung und
Forschung, the Heisenberg--Landau Programme (grant 2005), and
the Russian Foundation for Fundamental Research
(grants No.\ 03-02-16816, 03-02-04022 and 05-01-00992).

\begin{appendix}
\appendix

\section{QCD {\boldmath $\beta$} function at NLO}
\label{app:QCD-PT}

The first coefficients of the $\beta$ function are
\begin{eqnarray}
    b_0=\frac{11}{3}\,C_\text{A} - \frac{4}{3}\,T_\text{R} N_f
    \,,\qquad \qquad
    b_1=\frac{34}{3}\,C_{\text{A}}^{2}-
    \left(4C_\text{F} + \frac{20}{3}\,C_\text{A}\right)T_\text{R} N_f
    \,.
\label{eq:beta0&1}
\end{eqnarray}
Here, $T_\text{R}=1/2$ and $N_f$ denotes the number of flavors, whereas
the expansion of the $\beta$-function in the NLO approximation is given
by
\begin{equation}
 \beta(\alpha_{s}(\mu^2))
  = -\alpha_{s}(\mu^2)
      \left[b_0\left(\frac{\alpha_{s}(\mu^2)}{4 \pi}\right)
           + b_1\left(\frac{\alpha_{s}(\mu^2)}{4 \pi}\right)^2
      \right]\,.
\label{eq:betaf}
\end{equation}

\section{NLO correction to the pion form factor}
\label{app:NLO-pion-FF}

Here we present the detailed expressions for the color decomposition of
the NLO correction to the hard amplitude $T_\text{H}$, which describe
the factorized part of the pion form factor \cite{MNP98,BPSS04}
(see Eqs.\ (\ref{eq:THNLOpff})--(\ref{eq:TH12F})):
\begin{eqnarray}
 t_{\text{H},1}^{(1,\text{F})}(x,y)
  &=& \frac{N_\text{T}}{\x \y}
         \left[
               -\frac{28}{3}
               + \left( 6 -\frac{1}{x} \right) \ln \x
               + \left( 6 -\frac{1}{y} \right) \ln \y
               + \ln^2 (\x \y)
         \right]\,;
\label{eq:TH11F}
  \\[1.5mm]
 t_\text{H}^{(1,\text{G})}(x,y)
  &=& \frac{2N_\text{T}}{\x \y}
       \left[- \frac{10}{3}
             + \ln\left(\frac{\x}{x}\right)\ln\left(\frac{y}{\y}\right)
             - 4\left(\frac{\ln\x}{x}+\frac{\ln\y}{y}\right)
            \! - \! \tilde{H}(x,y) - R(x,y)
        \right] .  ~~~~~~
\label{eq:TH1G}
\end{eqnarray}
The functions $\tilde{H}(x,y)$ and $R(x,y)$ are defined by
\begin{eqnarray}
 \tilde{H}(x,y)
  = \left[\Li\left(\frac{\y}{x}\right)
           + \Li\left(\frac{\x}{y}\right)
           + \Li\left(\frac{x y}{\x\y}\right)
           - \Li\left(\frac{x}{\y}\right)
           - \Li\left(\frac{y}{\x}\right)
           - \Li\left(\frac{\x\y}{x y}\right)
         \right]~~~~
\label{eq:H}
\end{eqnarray}
and
\begin{eqnarray}
 R(x,y)
\!\! \!\!\! &&=\! \frac{1}{(x-y)^{2}}
       \Bigl[(2xy-x-y)(\ln x+\ln y)
        - (y\y^{2}+x\x^{2})(1-x-y)\tilde{H}(x,\y)
       \nonumber \\
  & &  
  - 2\left(xy^{2}+y^{2}-5xy+y+2x^{2}\right)
              \frac{\ln\y}{y}
    - 2\left(yx^{2}+x^{2}-5xy+x+2y^{2}\right)
              \frac{\ln\x}{x}
       \Bigr].
\label{eq:R}
\end{eqnarray}

\noindent
\section{``Analytization'' of powers of the coupling
         multiplied by logarithms}
\label{app:HSAnlo}

We present here the derivation of
${\cal L}_{2}^{(2)}\left(Q^2\right)$,
done in collaboration with S.\ Mikhailov.
To this end, let us first introduce
\begin{equation}
\label{eq:a_MS}
  a_s \left(Q^2\right)
   \equiv
    \frac{b_0}{4\pi}\,\alpha_{s}\left(Q^2\right)\,.
\end{equation}
For this quantity we can write a renormalization group solution in
the form
\begin{eqnarray}
\label{eq:a_MS_RG}
  \left[a_s^{(2)}\left(Q^2\right)\right]^2
   \ln\left(\frac{Q^2}{\Lambda^2}\right)
   = a_s^{(2)}\left(Q^2\right)
   + \left[a_s^{(2)}\left(Q^2\right)\right]^2\,
      c_1\,\ln\left[\frac{a_s^{(2)}\left(Q^2\right)}
                         {1+c_1a_s^{(2)}\left(Q^2\right)}
              \right]\,.
\end{eqnarray}
Expanding the expression $\ln[1+c_1a_s^{(2)}\left(Q^2\right)]$ and
retaining terms up to order $a_s^2$, we find
\begin{eqnarray}
\label{eq:a_MS_RG-2L}
  \left[a_s^{(2)}\left(Q^2\right)\right]^2
   \ln\left(\frac{Q^2}{\Lambda^2}\right)
   = a_s^{(2)}\left(Q^2\right)
   + \left[a_s^{(2)}\left(Q^2\right)\right]^2\,
      c_1\,\ln\left[a_s^{(2)}\left(Q^2\right)\right]\,.
\end{eqnarray}
To get rid of the logarithm, we use the following trick
\begin{eqnarray}
\label{eq:a_MS_RG-2L_der}
  \left[a_s^{(2)}\left(Q^2\right)\right]^2
   \ln\left(\frac{Q^2}{\Lambda^2}\right)
   = a_s^{(2)}\left(Q^2\right)
   +  c_1\,
      \frac{d}{d\varepsilon}\left[a_s^{(2)}\left(Q^2\right)
                            \right]^{2+\varepsilon}
       \Big|_{\varepsilon=0}
\end{eqnarray}
and return to the original coupling to obtain
\begin{eqnarray}
\label{eq:alpha_s_RG-2L}
  \left[\alpha_s^{(2)}\left(Q^2\right)\right]^2
   \ln\left(\frac{Q^2}{\Lambda^2}\right)
   = \frac{4\pi}{b_0}\,\alpha_s^{(2)}\left(Q^2\right)
   +  c_1\,\frac{4\pi}{b_0}\,
      \frac{d}{d\varepsilon}\left[\alpha_s^{(2)}\left(Q^2\right)
                            \right]^{2+\varepsilon}
       \Big|_{\varepsilon=0}\,.
\end{eqnarray}
Now we can proceed with the ``analytization '' of the term
$[\alpha_s^{(2)}\left(Q^2\right)]^2\ln\left(Q^2\right)$,
giving rise to analytic expressions for non-integer powers of the
coupling, i.e.,
\begin{eqnarray}
\label{eq:alpha_s_RG-2L_APT}
  \left\{\left[\alpha_s^{(2)}\left(Q^2\right)\right]^2
   \ln\left(\frac{Q^2}{\Lambda^2}\right)\right\}_\text{an}
   = \frac{4\pi}{b_0}\,{\cal A}_1^{(2)}\left(Q^2\right)
   +  c_1
      \left[\frac{d}{d\varepsilon}\,
       {\cal A}_{2+\varepsilon}^{(2)}\left(Q^2\right)
      \right]_{\varepsilon=0}\,.
\end{eqnarray}
Using the representation~\cite{BMS05}
\begin{eqnarray}
\label{eq:A_nu+1_APT}
  \left(\frac{b_0}{4\pi}\right)^2{\cal A}_{\nu}^{(2)}\left(Q^2\right)
  = \frac{-1}{\Gamma(\nu)}\,
     \sum_{n\geq0}
      \zeta(1-\nu-n)
       \frac{\left[-\ln\left(Q^2/\Lambda^2\right)\right]^n}
            {\Gamma(n+1)}
\end{eqnarray}
and performing the differentiation,
we finally obtain
\begin{eqnarray}
  \left\{\left[\alpha_s^{(2)}\left(Q^2\right)\right]^2
   \ln\left(\frac{Q^2}{\Lambda^2}\right)\right\}_\text{an}
   = \frac{4\pi}{b_0}\,
      \left[{\cal A}_1^{(2)}\left(Q^2\right)
          + c_1\,\frac{4\pi}{b_0}\,
             f_{\cal L}\left(Q^2\right)
      \right]\,,
\end{eqnarray}
with $f_{\cal L}\left(Q^2\right)$ being defined in Eq.\ (\ref{eq:f_MS}).
\end{appendix}



\end{document}